# Sculpting the shape of semiconductor heteroepitaxial islands: from dots to rods

J.T. Robinson[1,2], D.A. Walko[3], D.A. Arms[3], D.S. Tinberg[4], P.G. Evans[4], Y. Cao[1,2], J.A. Liddle[2], A. Rastelli[5], O.G. Schmidt[5,6], and O.D. Dubon[1,2]

[1]*Department of Materials Science and Engineering, University of California, Berkeley, CA 94720*

[2]*Materials Sciences Division, Lawrence Berkeley National Laboratory, Berkeley, CA 94720*

[3]*Advanced Photon Source, Argonne National Laboratory, Argonne, IL 60439*

[4]*Department of Materials Science and Engineering, University of Wisconsin, Madison, WI 53706*

[5]*Max-Planck-Institut für Festkörperforschung, Heisenbergstrasse 1, D-70569 Stuttgart, Germany*

[6]*Institute for Integrative Nanoscience, IFW Dresden, Helmholtzstrasse 20, D-01069 Dresden, Germany*

(Dated: January 31, 2007)

## ABSTRACT

In the Ge on Si model heteroepitaxial system, metal patterns on the silicon surface provide unprecedented control over the morphology of highly ordered Ge islands. Island shape including nanorods and truncated pyramids is set by the metal species and substrate orientation. Analysis of island faceting elucidates the prominent role of the metal in promoting growth of preferred facet orientations while investigations of island composition and structure reveal the importance of Si-Ge intermixing in island evolution. These effects reflect a remarkable combination of metal-mediated growth phenomena that may be exploited to tailor the functionality of island arrays in heteroepitaxial systems.

Surface layers of metal can greatly influence the growth behavior of epitaxial semiconductor structures.  For example, metal overlayers have been shown to mediate the growth of high-quality, single-crystalline films in systems in which island growth or poor epitaxy would otherwise occur[1-4].  They have in addition been used to tune the characteristics of epitaxial islands, or quantum dots, including island size, density and to a limited extent shape[5,6]. These and other related effects of metal overlayers on growth morphology have become known as the surfactant effect[1] (although the term "surfactant" is not applied in the strictest definition of the word in all cases). In this Letter we show that by patterning metal overlayers on the Si surface, one can controllably and radically modify the shape of highly ordered Ge islands.

Growth of Ge on Si has been the model system for studying epitaxial semiconductor quantum dot assembly. It occurs via the Stranski-Krastanov (SK) mode, which consists of the formation of a Ge wetting layer followed by random island growth.  In general, investigations of this system have focused on the characterization of fundamental aspects of island evolution, structure, and composition[7-11] as well as the realization of routes for island array assembly[12-15].  While some success has been achieved in producing island assemblies, up to now islands have been largely restricted to certain shapes such as the widely observed huts, domes and superdomes in the case of Ge grown on Si (001).

On Au-patterned Si (001), deposited Ge atoms assemble into an extensive, highly ordered array of hundreds of thousands of islands, limited by the extent of the pattern rather than stochastic effects.  The process involves a few simple steps described elsewhere[16].  Briefly, a Si wafer is rinsed in methanol, H-terminated in a dilute HF:$H_2O$ (1:10) solution, and loaded into an electron-beam evaporation chamber for metal deposition. One nanometer of Au (or other metal) is deposited through a stencil mask, which contains arrays of square windows 200 nm on a side and is placed in direct contact with the Si wafer. After metal patterning, the stencil mask is removed, and the wafer is transferred to a molecular beam epitaxy reactor where Ge deposition is carried out at a substrate temperature of 600 ºC and a rate of ~9 ML/min. (1 ML of Ge = $6.27 \times 10^{14}$ cm$^{-2}$) using an elemental Ge effusion source. As a result of this process, islands grow at the center sites of the square lattice formed by the Au dots and thus themselves form a square lattice displaced from the Au-dot lattice.

Figures 1(a), (b), and (c) present atomic force microscopy (AFM) phase images of islands formed by Ge deposition on Au-patterned Si (110), (001), and (111), respectively. On Si (001) and (110) islands are principally bound by {111} facets, which lead to square-based truncated pyramids (TPs)[16] and rectangular-based nanorods, respectively. On Si (111) the islands instead form {113} facets establishing their approximately tetrahedral shape. These islands are strikingly different from those found on the clean, Au-free regions of each sample (Figs. 1d-f).

Growth of islands on Sn- and Ag-patterned Si produces similar ordering behavior to that on Au-patterned Si. However, island shapes are markedly different. To elucidate these differences, island shapes have been analyzed by transforming AFM height images into slope $|n|$ images, where $n = \nabla f(x, y)$ is the surface gradient and $f(x, y)$ is the surface height at position $(x, y)$[17,18]. In order to precisely identify the facets that bound a particular island geometry, we plot in a two-dimensional histogram the frequency at which the values of *n* appear in AFM images. Consequently, all the points associated with a given surface orientation contribute to the same spot in the resulting intensity plot, also referred to as a facet plot (FP)[18].

Figures 2 shows the grayscale slope images and corresponding FPs of Ge islands grown on metal-free and metal-patterned Si (110). The FPs represent statistics taken from about 25 islands for each sample. From the FP in Fig. 2(b), we find that the dome-like islands on metal-free Si

(110) contain two {105}, four {113}, and two {111}-type facets at the island base and four {15 3 23}-type facets at the island top. The faint lines connecting facet spots arise from the facet boundaries on each island. On Au-patterned Si (110), nanorods are bounded principally by {111} facets resulting in intense spots in the corresponding FP (Fig. 2d). Interestingly, analysis of nanorods at the earlier stages of growth show that the first facets to develop are of the {111}-type (FP inset in Fig. 2c). In comparison, the islands on Sn-patterned Si (110) have a rounded shape (Fig. 2e). Even so, the FP in Fig. 2(f) reveals these islands have the same facets as islands in the metal-free region, minus the {105}. Finally, islands grown on Ag-patterned Si (110) (Fig. 2g) have a distinctly different shape from those on Au- or Sn-patterned Si (110). In this case the FP (Fig. 2h) reveals that these islands are bound mainly by {113}-type facets.

The FPs in Figure 2 indicate that no "new" facets appear on islands in the metal-patterned regions. Instead we find enhanced growth of specific facets already present, producing islands with very different shapes. Gold greatly enhances the formation of {111} facets while Sn has a much weaker effect on promoting or inhibiting the growth of particular facets. In the case of the Ag-patterned Si (110), the formation of {113} facets is favored resulting in a noticeably different island from those on the Au- and Sn-patterned surfaces. Facet analysis of islands on the Si (001) surface yields equivalent results. Thus, we conclude that the metal species favors the formation of particular facets that give rise to the observed shapes. Decoration of the surface by a metal species may lead to marked differences in surface tensions, adatom diffusion, and even intermixing between deposited Ge and Si substrate atoms that influence island shape evolution. Recent *in situ* x-ray photoemission electron microscopy experiments indeed show that, while Au is immobile prior to Ge growth due to an oxide formed in the immediate vicinity of each patterned Au square, it is distributed over truncated pyramidal islands on Au-patterned Si (001).

For islands on Au-patterned Si (001) and (110), the formation of steep {111} facets at low coverages[16] (~3-4ML-Ge) is surprising. We have employed chemical etching to understand the processes by which these island structures are assembled. Wet chemical etching with a 30% solution of hydrogen peroxide ($H_2O_2$) selectively removes $Si_{1-x}Ge_x$ alloys with compositions having $x>0.65$ (i.e., Ge-rich material)[19]. The same islands were imaged via AFM *before* and *after* $H_2O_2$ treatment, providing a quantitative measure of the volume of material removed. Figures 3(a-d) show three-dimensional AFM images of two islands analyzed before and after etching. The TP in Figs. 3(a) and (b) was formed at a nominal Ge coverage of 5 ML. After etching, the TP shows only a slight change of volume indicating that most of it contains Si to a level of at least 35%, or $Si_{0.35}Ge_{0.65}$. Indeed, small lens-like island structures from which TPs evolve displayed *no* measurable change in island volume upon exposure to $H_2O_2$. In islands produced at the higher Ge coverage of 10 ML, significant mass is removed by peroxide etching as revealed in Figs. 3 (c) and (d). Etching of Ge-rich material from islands of this and larger sizes exposes a highly Si-Ge intermixed core that reflects the truncated pyramidal shape of the island at the earlier stages of growth. The presence of a Si-Ge intermixed core occurs irrespective of island shape as reflected in Figs. 3(e-f), which present AFM images of nanorods after etching; at low deposited Ge coverages peroxide etching leads to little change in the post etched volume (Fig. 3e) while at higher coverages an intermixed core along the nanorod length is uncovered (Fig. 3f).

Figure 3(g) shows a plot of the change in volume upon etching versus the original island volume. A slope close to zero associated with islands of small volume indicates little change in volume upon etching and thus extensive intermixing throughout the structure. Large islands are described by a slope close to unity indicating that they are growing with an alloyed core of constant volume and increase in size by the addition of Ge-rich material to the outside. This behavior is observed for islands grown on both Au-patterned Si (001) and (110). The Si-Ge intermixing observed at low coverages in metal-patterned silicon differs from the Ge islanding behavior reported on metal-free Si (001). In the latter case, a fluctuation in surface flatness that typically leads to the formation of {105} facets does not provide a suitable path for the formation of {111} facets. Indeed, similar etching analysis of {105} faceted huts reveal these islands have Si-rich edges/corners and *Ge-rich cores*[19]. Evolution of huts into domes, which are composed of steeper {113} and {15 3 23} facets, results in the transition to a more Si-rich core and Ge-rich shell[20]. Thus, we believe in metal-patterned silicon enhanced Si-Ge intermixing provides a path for the collection of sufficient volume within an island by a ripening process to support the formation of {111} facets at a relatively early stage of island formation.

In order to further quantify the structural and compositional properties of ordered islands arrays, we have analyzed Au-patterned Si samples with a variety of Ge coverages using x-ray microdiffraction. Measurements were performed on beamline 7ID of the Advanced Photon Source (APS). X-rays were monochromatized to an energy of 10.5 keV and focused to a spot size of 15x15 $\mu m^2$. Reciprocal-space maps (RSMs) in high-symmetry planes were collected near several bulk Si diffraction peaks. Figure 4 shows a series of RSMs near the Si (111) Bragg peak for three samples with nominal Ge coverages of 6 ML, 9 ML and 50 ML on Au-patterned Si (001). At the lowest Ge coverage (Fig. 4a), the broad diffraction features from the TPs lie away from the high-symmetry line, indicating a strained, noncubic structure. Using continuum elasticity theory to correlate the in-plane and out-of-plane diffraction positions, we have determined that islands have a non-uniform composition (resulting in a radial spread in the diffraction peak) and are non-uniformly strained (resulting in the azimuthal spread). Although the islands are dislocated at this coverage, some of the epitaxial strain remains– up to about 25% of the total strain possible due to the lattice mismatch between Si and Ge. The average Ge composition of these TPs is x=0.65 although the radial breadth of the peak indicates that x varies from 0.8 to 0.4. Nanorod islands on Si (110) show a similar extent of intermixing. Our results for surfaces without the metal pattern are similar to those of Stangl *et al*.[21] for a sample grown under comparable conditions.

Germanium islands formed with 9 ML and 50 ML of Ge have four distinct orientations related by rotations of 0.35° and 0.7º, respectively, from the surface normal. The two lobes in the reciprocal-space map of Figs. 4(b) and (c), which are symmetric about the high-symmetry (cubic) line, are signatures of this orientational splitting; the lobes due to the other two orientations lie out of the (*HHL*) plane shown in this figure. These islands are free of tetragonal distortion and are characterized by an average Ge concentration of x=0.77 and 0.95, respectively.

The relationship between intermixing and relaxation is most clearly revealed in the etching and microdiffraction experiments performed on Au-patterned Si (001) samples with nominally 8-9 ML of Ge. In this range of Ge coverage, islands have developed Ge-rich caps, which contain additional facets that give rise to the superdome-like shape, while reciprocal space maps show these islands to have become almost fully relaxed. We interpret this convergence of effects to indicate that intermixing ceases with the occurrence of nearly complete island relaxation and that island growth proceeds by the addition of Ge. These combined observations yield a picture for the evolution of islands on metal-patterned Si in which Si-Ge intermixing at the early stages of growth provides a path for the formation of specific steep facets as dictated by the choice of metal species and substrate orientation.

O.D.D. acknowledges support from the National Science Foundation under contract number DMR-0349257. The work at the Lawrence Berkeley National Laboratory was supported in part by the Director, Office of Science, Office of Basic Energy Sciences, Division of Materials Sciences and Engineering, of the U.S. Department of Energy under Contract No. DE-AC02-05CH11231. Experiments at the Advanced Photon Source were supported by the U. S. Department of Energy, Office of Science, Office of Basic Energy Sciences, under Contract No. DE-AC02-06CH11357.

**Figure Captions**

**Figure 1 -** AFM phase images show Ge islands grown on: (a) Au-patterned Si (110), (b) Au-patterned Si (001), (c) Au-patterned Si (111), (d) Si (110), (e) Si (001), and (f) Si (111) (image sizes: 750x750nm$^2$). Nominal Ge coverages are: (a) and (d) 7.5 ML; (b) and (e) 4 ML; (c) and (f) 5 ML. The principle facets bounding islands on Au-patterned Si are labeled in (a-c). Island orientation is dependent *only* on substrate orientation and not the orientation of the evaporated metal pattern relative to the substrate.

**Figure 2** - Grayscale slope images (a), (c), (e), and (g) and their corresponding facet plots (FPs) (b), (d), (f), and (h) of Ge islands grown on metal-free and metal-patterned Si (110) with approximately 8 ML-Ge (slope image scale: 500x500nm$^2$; facet plot scale: 1.8). (a) Islands on metal-free Si (110) are composed of {105}, {15 3 23}, {113}, and {111} facets, as determined from the FP in (b). (c) Islands on Au-patterned Si (110) are bounded by these same facets, although the {111}-type facets form first as shown by the FP inset (corresponding to 4 ML Ge) and compose the largest area of the nanorod island. (e) Islands on Sn-patterned Si (110) have a much more rounded shape and the FP in (f) shows that these islands have facets of the {113}, {15 3 23}, and {111} types. (g) Islands on Ag-patterned Si (110) are composed mainly of {113} facets.

**Figure 3** – (a-f) Perspective view of AFM height images of islands formed on Au-patterned Si (001) and (110) before and after peroxide etching (image sizes: 450x450 nm$^2$). (a) and (c) correspond to as-grown islands with 5 ML and 10 ML of Ge, respectively; island heights are approximately 36 nm and 48 nm, respectively (z=50 nm). (b) and (d) show the same islands in

(a) and (c) after $H_2O_2$ etching; islands heights are approximately 35 nm and 37 nm, respectively (z=50 nm). Islands formed by the deposition of (e) 4.5 ML and (f) 10 ML of Ge on Au-patterned Si (110) after peroxide etching (z=25 nm). Dotted lines have been included to mark the footprint of islands prior to etching. (g) Island volume change after $H_2O_2$ etching versus original island volume for islands grown on Au-patterned Si (001) (triangles) and Si (110) (squares). Having a slope of one and passing through the origin, the dashed line represents the case for complete island removal by peroxide etching. Volume measurements were performed on samples representing a range of Ge coverages.

**Figure 4** - Reciprocal-space maps from islands formed on Au-patterned Si (001) with Au dot spacing of 800 nm. This spacing was chosen because it produced a high level of island ordering and uniformity in island shape and size. These maps were collected via x-ray microdiffraction near the Si (111) peak for Ge coverages of (**a**) 6 ML, (**b**) 9 ML, and (**c**) 50 ML. The maps are in the *H=K* plane, with *L* being the direction of the surface normal (reciprocal-space units are referenced to the Si lattice). Contours are spaced logarithmically. The dashed lines represent the position of diffraction peaks due to unstrained $Si_{1-x}Ge_x$ terminating at the position of a peak due to pure Ge.

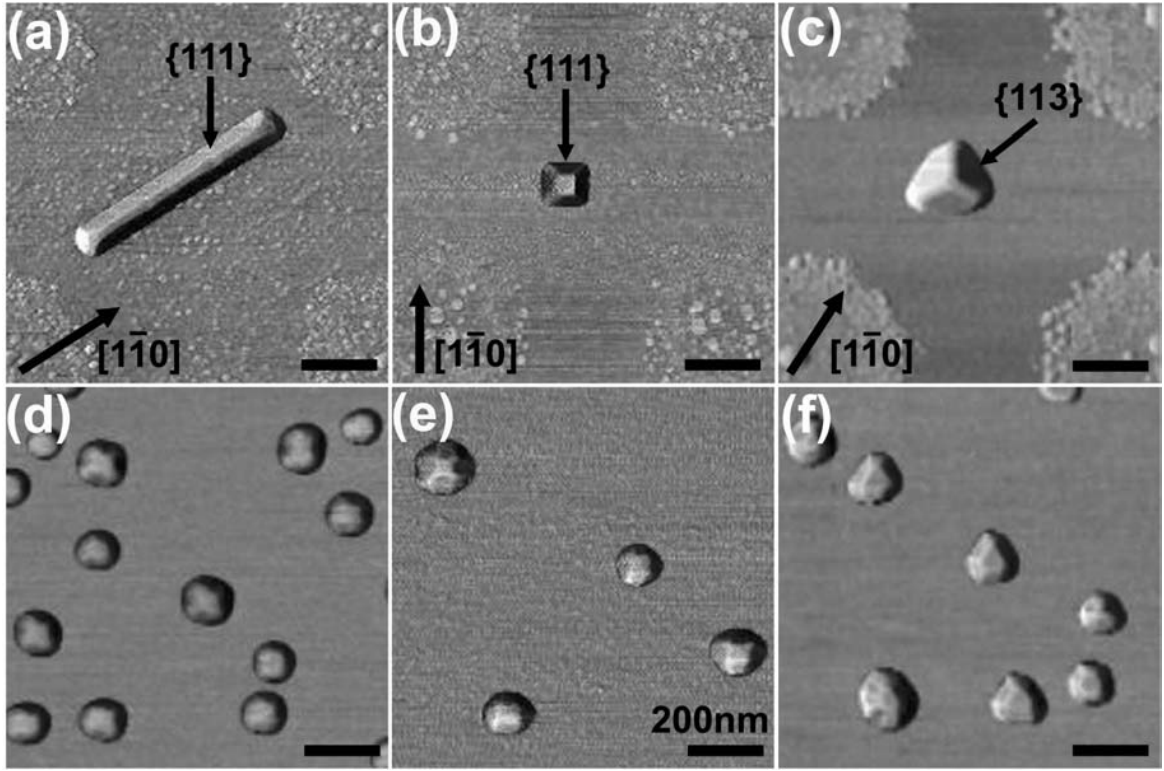

**Figure 1** - Robinson *et al.*

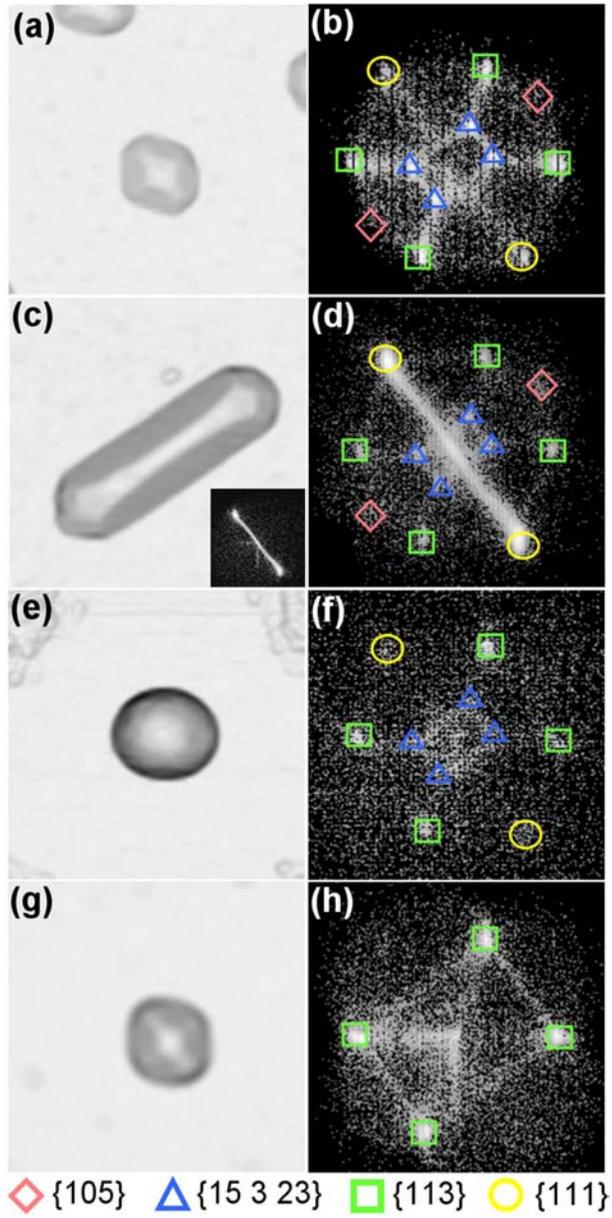

**Figure 2** - Robinson *et al.*

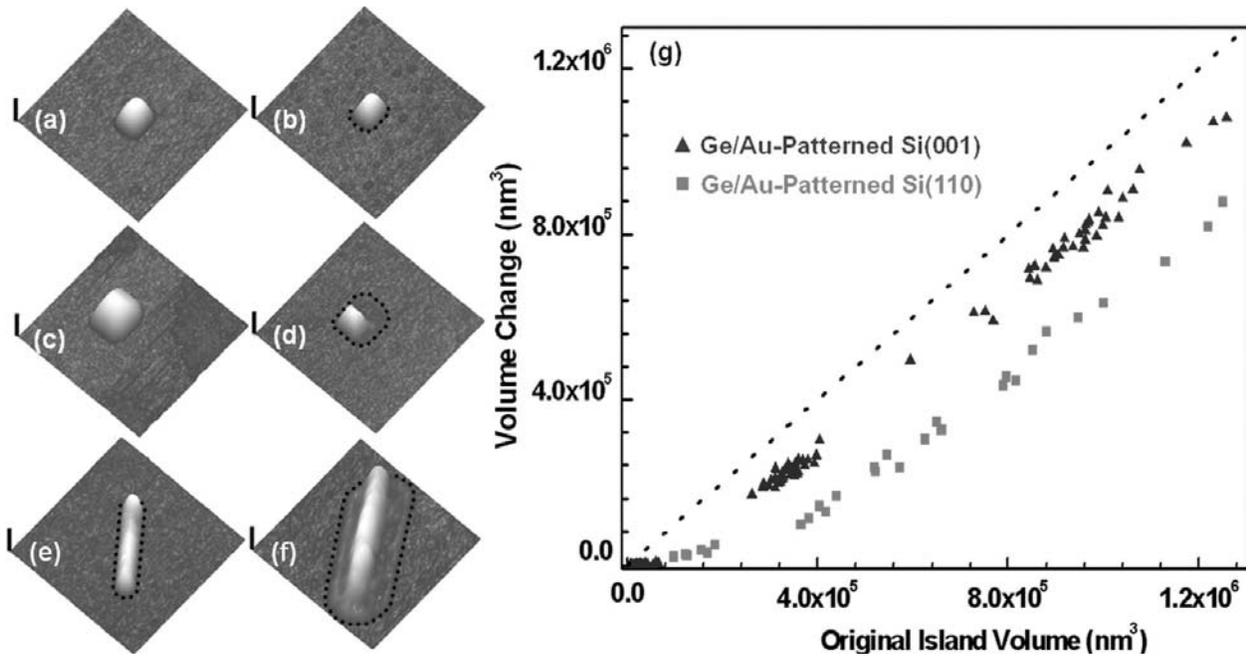

**Figure 3** - Robinson et al.

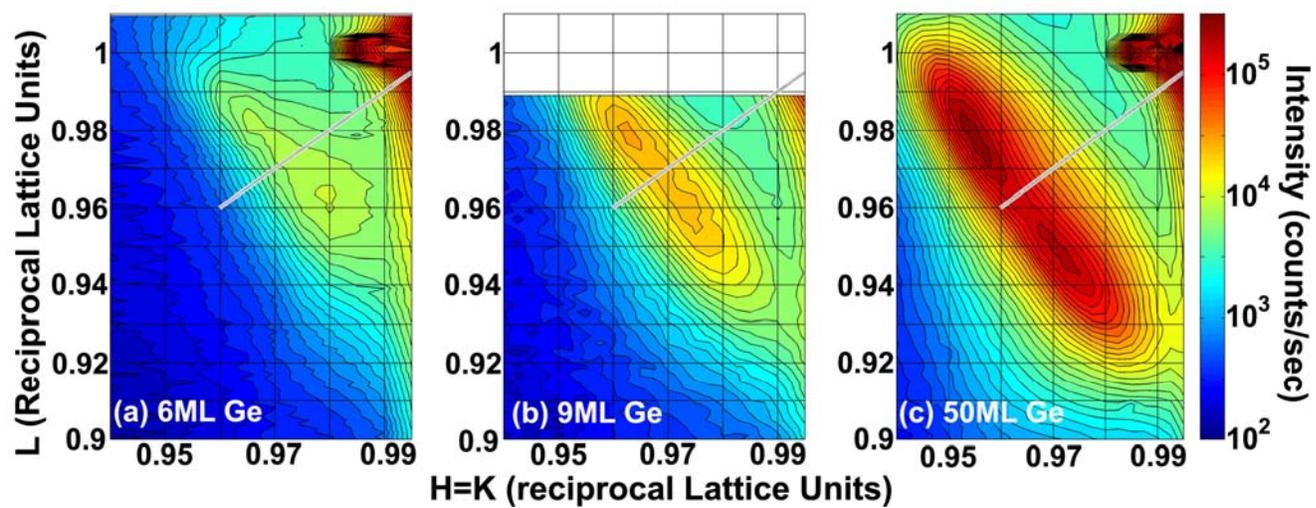

**Figure 4** - Robinson, *et al.*